\documentclass[]{aa}

\usepackage{amssymb}
\usepackage{latexsym}
\usepackage{textcomp}
\usepackage{array}
\usepackage{graphicx}
\usepackage{natbib}
\bibpunct{(}{)}{;}{a}{}{,}    
\usepackage[varg]{txfonts}
\usepackage{color}

\begin{document}

\newcommand{\revis}[1]{#1}

\title{Multilayer modeling of the aureole photometry during the Venus transit: comparison between \revis{SDO/HMI} and VEx/SOIR data
}
\titlerunning{2012 Venus transit: aureole photometry, SDO/HMI and VEx/SOIR}

\author{C. Pere  \inst{1,2}\fnmsep\thanks{PhD funded by the EU FP7 EuroVenus Project (G.A. $\#$606798).}, P. Tanga\inst{1} \and Th. Widemann\inst{2,3} \and Ph. Bendjoya\inst{1} \and A. Mahieux\inst{4,5} \and V. Wilquet\inst{4} \and A.C. Vandaele\inst{4}}
\offprints{C. Pere, \email{cpere@oca.eu}}
\institute{Laboratoire Lagrange, Université C\^ote d'Azur, Observatoire de la C\^ote d'Azur, CNRS, Blvd de l'Observatoire, CS 34229, 06304 Nice cedex 4, France  
              \email{cpere@oca.eu}
\and LESIA, UMR CNRS 8109, Paris Observatory, France
\and DYPAC, EA 2449, Universit\'e de Versailles-Saint-Quentin-en-Yvelines, Guyancourt,France
\and Belgian Institute for Space Aeronomy, 3 av. Circulaire, 1180 Brussels, Belgium
\and Fonds National de la Recherche Scientifique, Brussels, Belgium
}

\date{\today}

\abstract
{The mesosphere of Venus is a critical range of altitudes in which complex temperature variability has been extensively studied \revis{by the space mission Venus Express (VEx) during its eight-year mission} (2006-2014). In particular, the Solar Occultation in the InfraRed (SOIR) instrument probed the morning and evening terminator in the 70–170 km altitude region, at latitudes extending from pole to pole, using spectroscopic multiband observations collected during occultations of the Sun at the limb. Data collected at different epochs and latitudes show evidence of short and medium timescale variability as well as latitudinal differences. Spatial and temporal variability is also predicted in mesospheric and thermospheric terminator models with lower boundary conditions at 70 km near cloud tops. }
 {The Venus transit on June 5-6, 2012 was the first to occur with a spacecraft in orbit around Venus. It has been shown that sunlight refraction in the mesosphere of Venus is able to provide useful constraints on  mesospheric temperatures at the time of the transit. The European Space Agency’s Venus Express provided space-based observations of Venus during the transit. Simultaneously, the Venus aureole photometry was observed using ground-based facilities and solar telescopes \revis{orbiting}  Earth (NASA's \revis{Solar Dynamic Observatory}, JAXA's HINODE). \revis{As the properties of spatial and temporal variability of the mesosphere are still debated, the opportunity of observing it at all latitudes at the same time, offered by the transit, is rather unique. In this first paper, we establish new methods for analyzing the photometry of the so-called aureole that is produced by refraction of the solar light, and we investigate the choice of physical models that best reproduce the observations.}}
{We compared the refractivity profile obtained by SOIR at the time of the June 2012 transit to the aureole photometry. For this goal, we explored isothermal and multilayered refraction models of the terminator atmosphere based on the vertical density profile obtained by VeX/SOIR at a latitude of +49$^\circ$ and successfully compared it to the aureole photometry observed from space by the HMI instrument of the Solar Dynamic Observatory (SDO).} 
{We obtain an independent constraint of 4.8 ± 0.5 km for the aerosol scale height in the upper haze region above 80 km. We show that a full multiple-layer approach is required to adequately reproduce the aureole photometry, which appears to be sensitive to several second-order variations in the vertical refractivity. }
{}


\maketitle

\section*{Introduction}  
        
        A transit of Venus (ToV) in front of the Sun is a rare event, a unique opportunity to study the sunlight refraction in the atmosphere of the planet during ingress and egress, from
which the mesospheric and upper haze structure can be constrained.  
        The event of June 5-6, 2012 \revis{was the first transit in history to occur} while a spacecraft was in orbit around the planet\revis{, and was observable from a large portion of the Earth, stretching from Central-East Europe to the American continent, across the Pacific}. 
        
        Accounts of  past historic transits provided detailed descriptions of the \revis{planet morphology through telescopic observations} of Venus during ingress and egress phases that were relevant for contact timings \citep[for a review, see][]{Link-1969}. In the past, timings were the only scientific data collected, in the attempt to use the events for the determination of the solar parallax. \revis{While today this interest no longer exists, another category of phenomena, involving the atmosphere of Venus, appears to be more relevant and can be linked to a larger domain of investigations, including exoplanet transits \citep{Ehrenreich2012, Widemann-et-al-2012}.}
        
        The portion of the planetary disk that is outside the solar photosphere has been repeatedly perceived as outlined by a thin bright arc called the aureole. On June 8, 2004, fast photometry based on electronic imaging devices allowed the first quantitative analysis of the phenomenon \citep{Tanga-etal-2012}. 
        
        \revis{The accuracy of the observations in 2004 was limited
because the campaigns were not specifically organized to photometrically observe the aureole}, which was only confirmed at that time. Measurements in 2004 were essentially obtained using NASA’s then operating Transition Region and Coronal Explorer solar observatory (TRACE), the Tenerife Themis solar telescope, the Pic-du-Midi 50 cm refractor, and the DOT in La Palma (Spain) \citep{Pasachoff-etal-2011, Tanga-etal-2012}.
 Owing to the difficulty of reaching an acceptable signal-to-noise
ratio (S/N) next to the solar photosphere, a region that is typically contaminated by a strong background gradient, only the brightest portions of the aureole were sampled. This left a strong uncertainty on the faint end of the aureole evolution, when Venus is located
farther
away from the solar limb. In these conditions, it was not possible to probe the deepest refracting layers, which are close to tangential optical thickness $\tau =1$ of the Venus atmosphere.  An isothermal model was fitted to the usable data, which yielded a single value of the physical scale height for each latitude and an estimate of the vertical extension of the refracting layers that contribute to the aureole.
        
        On June 5-6, 2012, several observers used a variety of acquisition systems to image the event; these systems ranged from amateur-sized to professional telescopes and cameras. In
this way, a large amount of quantitative information on this atmospheric phenomenon was collected for the first time. For the 2012 campaign, initial results and observations have been presented \citep{Wilson-et-al-2012, Widemann-et-al-2012, Jaeggli-et-al-2013}. 
        \revis{Direct multiwavelength measurements of the apparent size of the Venus atmosphere were obtained by \citet{Reale-et-al-2015}. In addition, the Doppler shift of sub-millimeter  $^{12}$CO and $^{13}$CO absorption lines was mapped by \citet{Clancy-et-al-2015} using the James Clerk Maxwell Telescope (JCMT) to measure the Venus mesospheric winds at the time of the transit.}
        
\begin{figure}
\includegraphics[width=\linewidth]{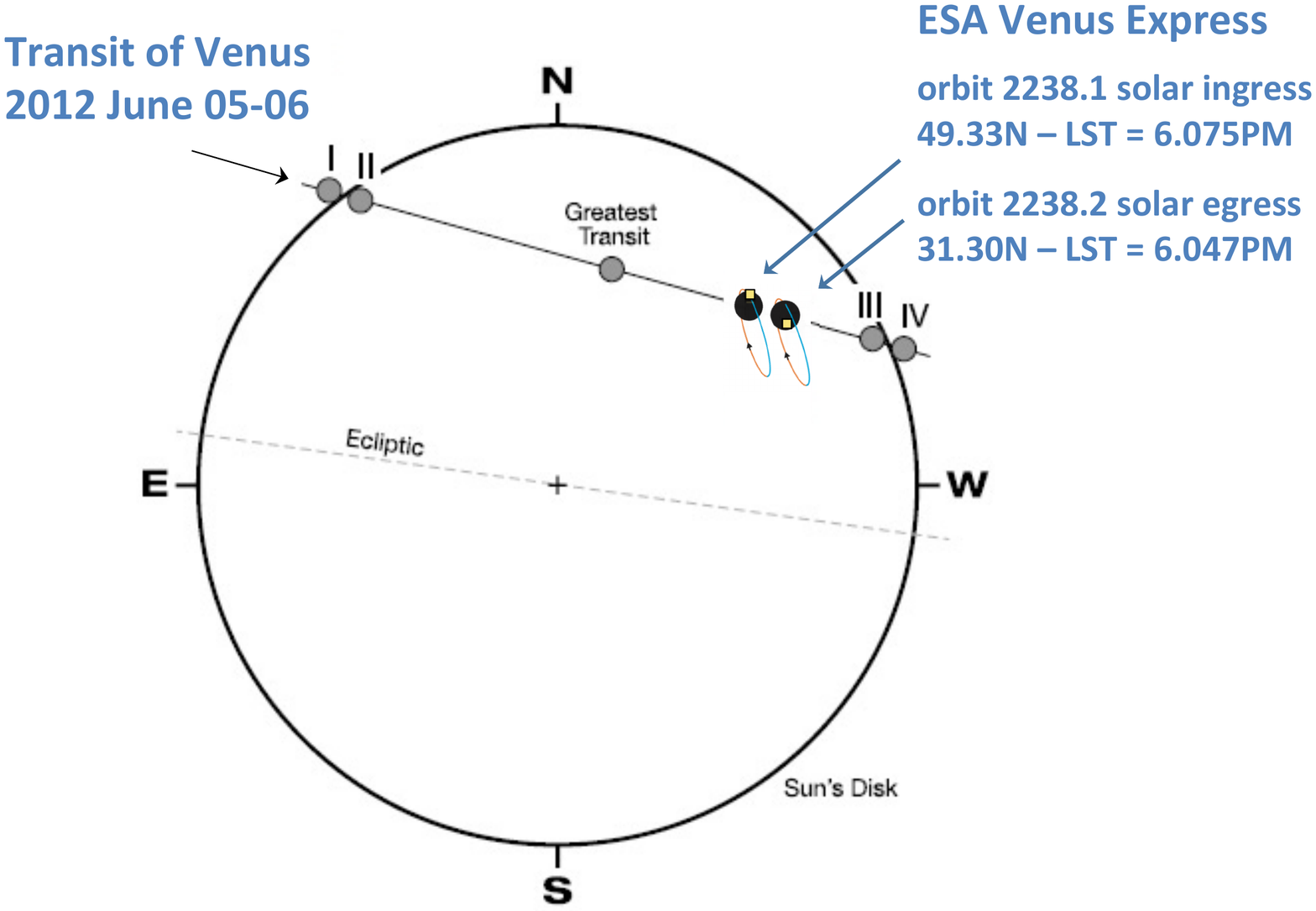}
\caption{Solar disk with the trajectory of Venus during the transit on June 5-6, 2012, as seen from the Earth geocenter. See the text for more details. }
\label{F:transit_scheme}
\end{figure}

        In this work, the first devoted to aureole photometry obtained during the June 2012 event, we use simultaneous data from the Earth-orbiting NASA Solar Dynamics Observatory (SDO) and Venus-orbiting ESA Venus Express spacecrafts.  Optical data retrieved from an image sequence of the Helioseismic and Magnetic Imager \citep[HMI, ][]{Schou-et-al-2012} \revis{onboard the SDO mission } are compared to atmospheric refraction models based on a vertical atmospheric density profile obtained by the VEx Solar Occultation in the Infrared \citep[SOIR, ][]{Vandaele-et-al-2008} instrument during orbit 2238. The Venus Express operations occurred while Venus was transiting the Sun, as seen from Earth. In Fig.~\ref{F:transit_scheme} \revis{the positions of the contacts (I to IV) are indicated, as well as the orbit of European Space Agency's Venus Express orbiter around the planet, projected to scale. Venus is shown at its location during  orbit 2238 when SOIR data were collected at the time of apparent solar ingress at latitude +49.33\textdegree on the evening terminator at 6.075 PM local solar time (LST). At the scale used in Fig. 1, the parallax effect from a site on the Earth surface, or from the position of SDO, is negligible. For data reduction, accurate positions of Venus relative to the solar limb are derived directly from SDO/HMI images.} 
        
        The advantage of SOIR is clearly related to the high vertical resolution obtained from its vantage observation point. From the Earth's ground-based and orbit-based telescopes such as SDO/HMI, the aureole vertical extension (corresponding to a few atmospheric scale heights) is unresolved. The aureole brightness is the result of the sum of refracted light at different altitudes. Although the evolution of the Earth-Venus-Sun geometry during the transit allows separating the contribution of different layers, \revis{both photometric and calibration accuracy limit the vertical resolution}. On the other hand, an advantage specific to transits is the possibility of simultaneously probing the entire limb of Venus, which allows deriving the atmospheric properties at all the latitudes where the aureole is observed.

In this paper, we test the applicability of an isothermal approach to SDO/HMI time-resolved photometry and the possible improvements provided by using a multilayer approach of the Venus atmosphere based on Venus Express data, in particular the SOIR vertical density profile obtained during the transit at a latitude of +49$^{\circ}$ to reproduce the aureole photometry. 
We also derive constraints on the upper haze altitude \revis{and on the} tangential opacity of the same latitude. 

The paper is organized as follows. First, we describe the condition of the 2012 ToV and the method for extracting and analyzing the data (Sect. 1). We then present three numerical models that we developed to study the aureole (Sect. 2), and we apply the different models to SOIR solar occultation data obtained at orbit \revis{2238 (Fig.\ref{F:transit_scheme}) to test their} consistency with aureole data (Sect. 3). 
        
\section{Observations by the Solar Dynamics Observatory}

\begin{figure}
\includegraphics[width=\linewidth]{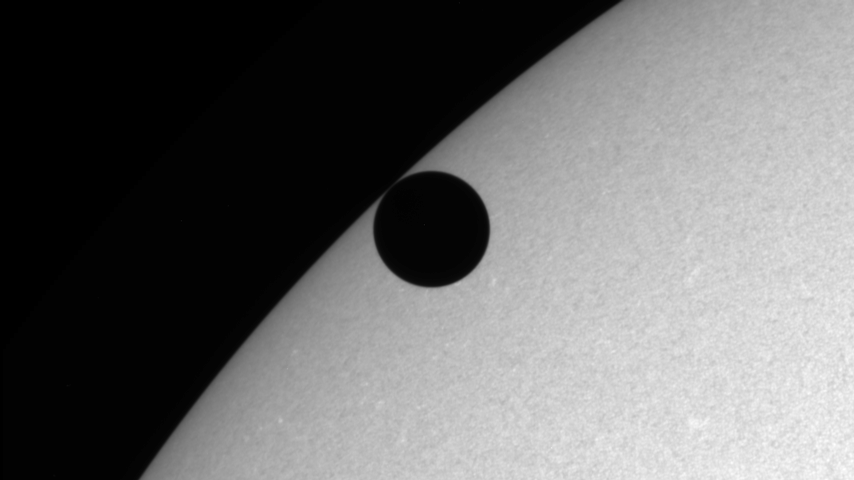}
\includegraphics[width=\linewidth]{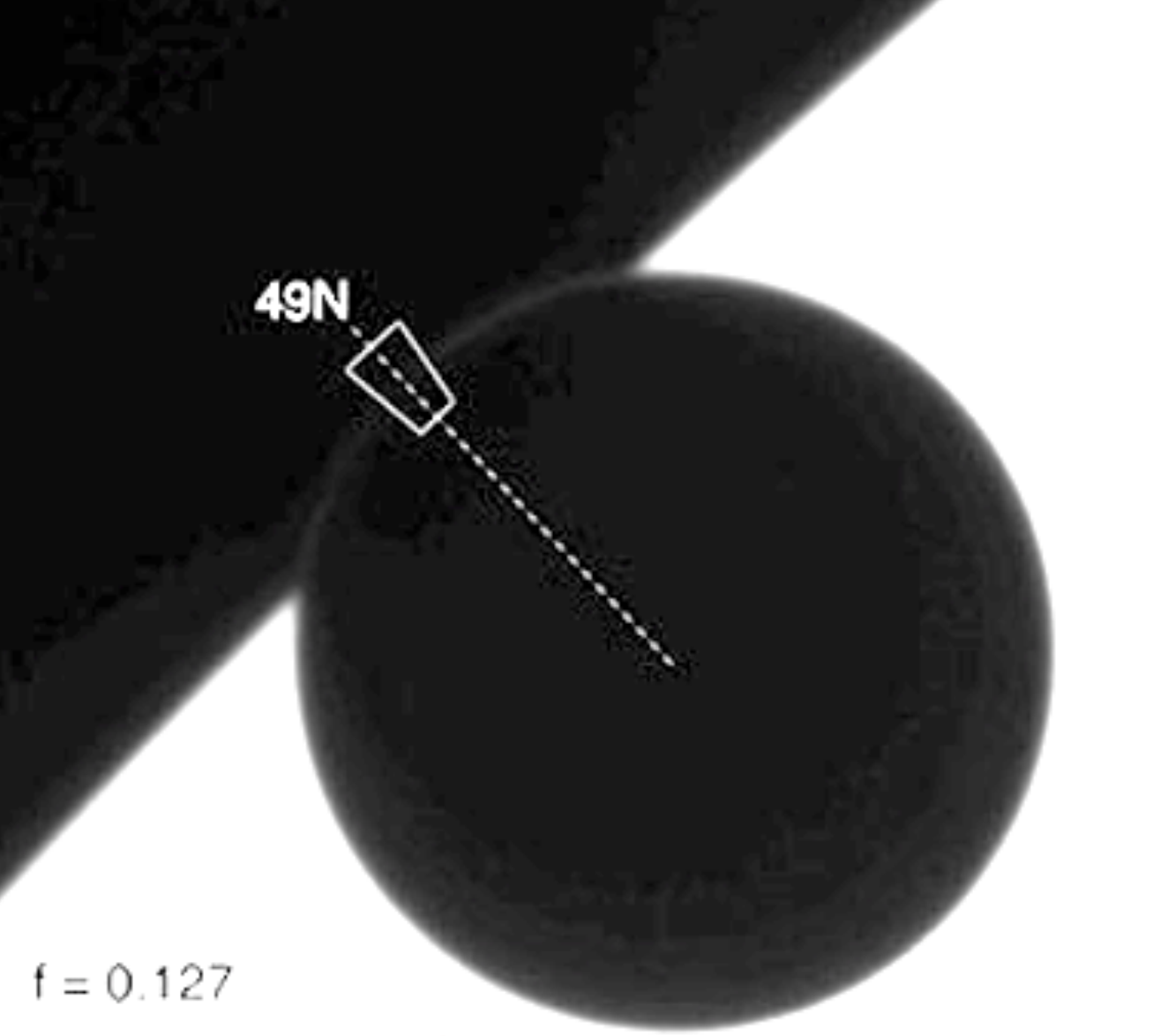}
\caption{Upper panel: Subframe centered on Venus \revis{at the epoch of} second contact, during the transit ingress, \revis{from a single SDO/HMI image obtained at 617.3 nm (Fe I absorption line)}. Lower panel: SDO image with extreme contrast stretch, to show the aureole. The radial direction of the flux measurement is shown, with the area at +49° that is considered for comparison to SOIR. \revis{The value of $f$ in the bottom left corner is the linear fraction of the Venus diameter projected outside the solar limb.}}
\label{Figure SDO}
\end{figure}

        The aureole photometry was derived from data acquired by the \textit{Helioseismic and Magnetic Imager} (HMI) instrument onboard the \textit{Solar Dynamics Observatory }(SDO, NASA),
which operates from an inclined geosynchronous orbit since 2010. A total of 776 images were obtained during the ingress of Venus and 862 during the egress, at a resolution of $\sim$0.504 arcsec per pixel, corresponding to 105 km at the distance of Venus at the transit epoch.   
HMI was designed to measure Doppler shift and magnetic field vector at the solar photosphere by exploiting the 617.3 nm Fe I absorption line. \revis{We exploited here the continuum images corresponding to Level-1.5 data products, implying that they have been normalized by flat-fielding but not rescaled or modified further.  During the transit, the time sampling interval is 45 seconds}. A  854×480 pixel subframe of a HMI image of is shown in Fig.~\ref{Figure SDO}. 

\begin{table}    
    \centering
    \begin{tabular}{l l l}
    \hline 
    Image sequence start (UT) & end (UT) & n. images \\
20:00:02.66 &  22:50:02.66 & 776 \\
    \hline 
    \multicolumn{3}{l}{Center wavelength: 617.3 nm } \\ 
    \hline  
    \multicolumn{3}{l}{Image scale: 0.504 arcsec/pixel} \\ 
    \hline  
    \multicolumn{3}{l}{Apparent Venus diameter (from geocenter): 57.80 arcsec  } \\ 
    \hline
    \end{tabular}
    \caption{Summary of the main properties of the transit observations by SDO (ingress only). The Venus diameter is derived from computing the planet ephemerides.}   
    \label{T:SDOprop}           
\end{table}






        
\subsection{Aureole brightness determination}
        
The photometry of the aureole consists of measuring the flux along a circular annulus containing the limb of Venus. The sector containing the aureole corresponds to the annulus portion projected against the background sky, outside the solar photosphere. 

\revis{To correctly determine the exact position of the aureole, we proceeded by fitting a circle to the limb of Venus, on two reference images where the planet is at least partially silhouetted against the Sun (with f<0.5). During the short duration of the ingress and egress, the motion of the planet relative to the Sun is essentially linear. Starting from the two reference positions, this allowed us to determine by extrapolation the position of Venus on all other images, which were previously aligned on the Sun.}

The measurement was repeated for each image to study the variation of the aureole brightness over time.
Our procedure started by extracting the transverse brightness profile of the aureole in the planetocentric radial direction. This was obtained by estimating the contribution of individual pixels in analog-to-digital units (ADU), using subpixel increments. At each step, a bilinear interpolation was applied to obtain the flux value at the corresponding position. 

The procedure, in absence of very steep brightness gradients, was sufficient to obtain a rather smooth profile. However, to further reduce the possibility that small fluctuations introduce noise on the curves by pixel-to-pixel variations, we averaged ten radial profiles spaced by 0.1\textdegree \ in latitude to obtain a final radial curve associated to a 1\textdegree \ interval.

\begin{figure}
\includegraphics[width=\linewidth]{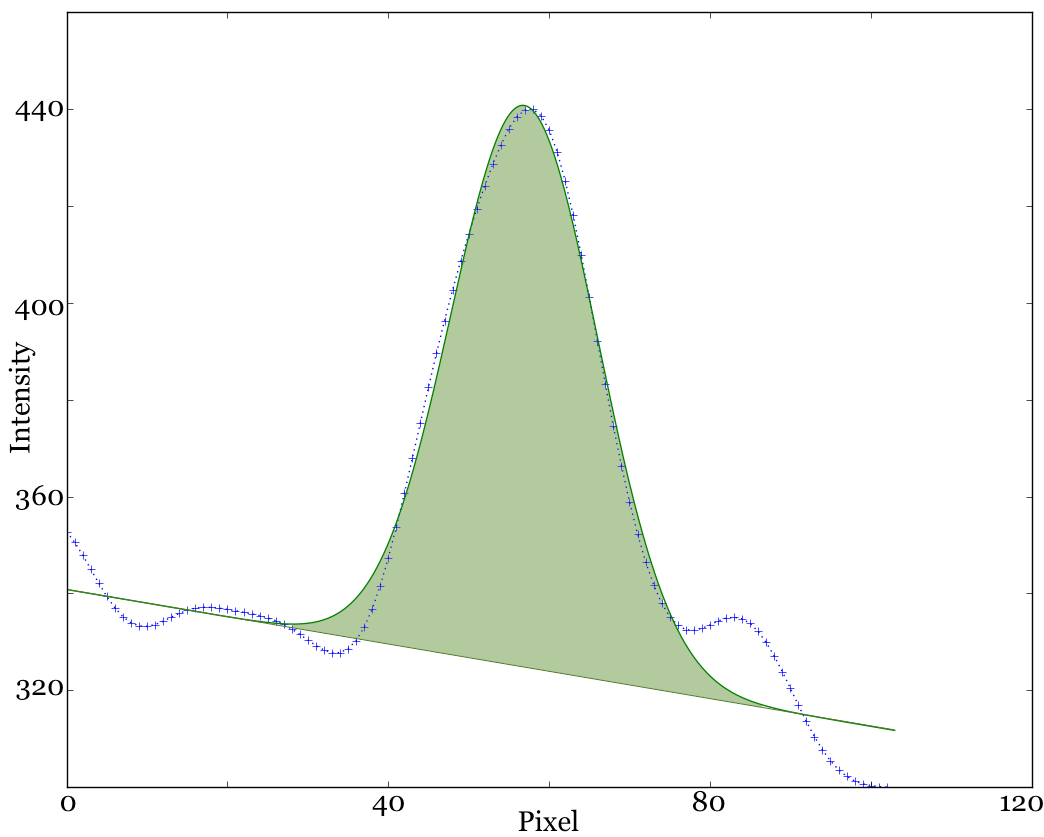}
\caption{\revis{Radial intensity profile of the aureole (blue crosses) as a function of the number of HMI pixels along the radial direction from the center of Venus. The profile has been measured from one SDO/HMI frame, collected on June 5th, 2012 at 22:21:55, for the latitude +49\textdegree. The blue dots represent the bilinear interpolation performed on the radial profile at steps of 1/10 of one pixel.} The vertical axis is the signal intensity in ADU. The green curve is the result of a Gaussian fit with a linear slope. The width at half-height is 1.875 arcsec and corresponds to 390 km at Venus.}
\label{Figure 3}
\end{figure}

The typical signal was well approximated by a Gaussian (an example is shown in Fig. \ref{Figure 3}), representing the transverse cut of the line spread function of the imaging system. \revis{At this stage, the position of the peak on the profile was verified to ensure that the planet position was correctly computed. The atmospheric scale-height of about 5~km is therefore unresolved by a factor $\approx$80.} As our measurements are performed very close to the Sun, a background signal fading away from the limb, mainly due to scattering in the telescope optics, is always present, and it can be modeled as a linear slope added to the Gaussian.  

We thus modeled the radial profile by a function $F_t$ as

\begin{equation}\label{eq:1}
\left\{
\begin{array}{ll}
        F_t(X) &= F(X)+F_b(X), \\
        F(X) &= g\ e^\frac{-(X-X_0)^2}{\sigma^2},\\
        F_b(X) &= a*X+b \end{array} \right.
,\end{equation}

where X is the radial position. The parameters \textit{a}, \textit{b}, \textit{g}, $X_0$ , and $\sigma$ were determined by a non-linear least squares fit on each of the profiles.

The integral of the Gaussian component over the width of a ring surrounding the aureole $\int {F(X)}$ represents the background--subtracted aureole flux (Eq.~\ref{eq:1}). This approach is different from the aperture photometry adopted by \citet{Tanga-etal-2012} and allowed us to better evaluate both the background and the aureole signal.

The aureole flux in ADU/pixel was converted into ADU per arcsec (i.e., the brightness of an aureole arc of 1~arcsec length)  and then normalized to the brightness of a 1~arcsec$^2$ of photosphere, measured at 1~Venus diameter from the solar limb.
        
With this method, the flux was measured at steps of 1\textdegree\
 in latitude.

\begin{figure}
\includegraphics[width=\linewidth]{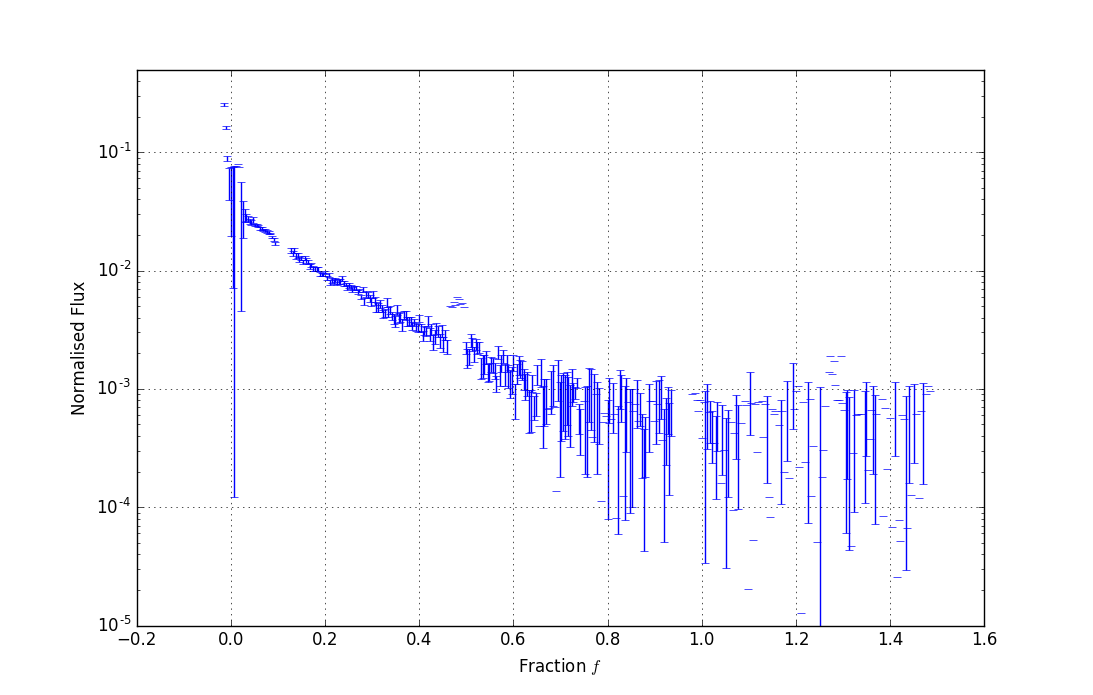}
\caption{\revis{Aureole flux at a latitude of +49\textdegree is plotted as a function of the fraction of the Venus diameter seen outside of the photosphere $f$.} The flux is normalized to a reference element of the solar photosphere, the brightness of a 1$\times$1~arcsec$^2$ at one apparent Venus diameter from the solar limb. The aureole flux is plotted as a function of the linear fraction of the Venus diameter seen outside of the photosphere $f$, from the point of view of the observer.}
\label{F:lightcurve}
\end{figure}
        
        Following \citet{Link-1969}, we define the ``phase'' of the Venus egress/ingress ( \textit{f} ) as the linear fraction of the planet diameter projected outside the solar limb (Eq.~\ref{eq:fraction}), as seen from a given observer. For instance, the value \textit{f}=0 corresponds to the planet disk entirely projected on the Sun and internally tangent to its limb; at \textit{f}=0.5 the center of the planet falls exactly on the limb of the Sun.
        
        Of course, $f$ is a function of time, but it can also be derived directly from the images by measuring the planet position relative to the solar limb. In practice, we extracted from the images the epoch of the first contact (when Venus is externally tangent to the limb,$t_{1st}$) and the second contact (when Venus is internally tangent to the limb, $t_{2nd}$). As the epoch of any image ($t_i$) is known, assuming a linear motion of the planet (appropriate over the $\sim$20~min duration of ingress or egress), the corresponding value of $f$ can easily be derived:
                
\begin{equation}
        f = - \frac{t_{i} - t_{2nd}}{t_{2nd} - t_{1st}}
\label{eq:fraction}
.\end{equation}

        By considering the whole image sequence obtained by SDO, we can then represent the phenomena observed with the evolving geometry of Venus with respect to the Sun and the observer by the parameter $f$, a proxy of time, which is directly related to the evolving geometric configuration.
        
        The position angle along the disk of Venus can easily be converted into a latitude by considering the known orientation of the SDO images (solar north up) and by computing the physical ephemerides of the Sun and Venus during the transit. At that epoch, the planetocentric latitude of the sub-Earth point was just 1\textdegree, implying that the discrepancy between position angle and latitude was very small and can be neglected in practice (1\textdegree\  at most at the poles).
        
        For a given latitude, that is, for a given point along the planet limb, our capability of observing the aureole is related to the interval of $f$ values for which that point is projected on the sky background, outside the solar photosphere. For this reason, the measurements of different latitudes span a different range of $f$. At the limit of large $f$ (planet largely outside the solar limb), the signal disappears into noise for any latitude that is considered. This occurs at levels $\sim$10$^{-4}$ of normalized flux. 
        
        At small $f$, the geometric limit is represented by the position at which a given point on the planet limb touches the solar limb. However, since we integrated the flux of the aureole radially over ten pixels to estimate the background, a practical limit exists and is reached sooner than the geometric limit. It corresponds to the contamination by the photosphere margin, which directly enters the measured annulus. For this reason, we conservatively removed the extreme of the curve at the limit of the smallest $f$ ($\approx$0.02) where a discontinuity in the flux indicates that the photosphere contaminates our measurements.
        
        Given the high rate of sampling, we additionally averaged the aureole flux over bins of ten single measurements. We then computed error bars from the standard deviations within each bin.
        
Figure~\ref{F:lightcurve} shows the light curve extracted at latitude +49\textdegree during the ingress at the morning terminator. Error bars include the contribution of photon noise from the background, the source, and the photospheric comparison. As expected, the curve presents an exponential decrease in brightness with increasing $f$, that is, at larger distances between the disk of Venus and the solar limb. By considering the light-curve section where a trend is clearly visible above the noise level, we are able to trace the aureole brightness over two orders of magnitude.

The maximum brightness of the aureole (around 10$^{-1}$, normalized units) occurs at very low $f$ values. At this geometry the luminosity is dominated by sunlight crossing the atmosphere at the highest altitude probed by the aureole. The corresponding sunlight beams are affected by a very small (subarcsec) total deviation that
is due to refraction. As the aureole image formed by refraction preserves the surface brightness of the source (in case of a perfectly transparent atmosphere), our normalization implies that the thickness of the atmospheric layer contributing to the aureole is $\sim$10$^{-1}$ arcsec$=20$~km. If additional opacity due to light scattering by aerosol particles above the cloud tops is included, the real altitude range can be higher. 

To derive physical parameters of the mesosphere where the refraction occurs, we try as a first approximation to adopt a model of transparent, isothermal atmosphere along the line of the initial model developed by \citet{Tanga-etal-2012}. As shown below, the light curve that we observe induced us to refine the isothermal assumption and adopt a multilayered approach for the refraction model. 

\section{Sunlight refraction models}

\subsection{Isothermal model (model 1)}

At first order, the aureole of Venus can be reproduced by a model taking into account the refraction of a finite array of elementary light sources originating from the solar photosphere. Our first approach is the isothermal model used in \citet{Tanga-etal-2012} for the interpretation of the transit data collected in 2004 by ground-based telescopes. The core of this model, called model 1 in the following, is based on the hypothesis of a transparent atmosphere as presented by \citet{Baum-Code-1953}. We recall its main properties below.

The refraction angle $\omega$ of a light ray that crosses the atmosphere and reaches the observer is given by 
        
\begin{equation}
\omega = -\nu(r)\sqrt{\left( \frac{2 \pi r}{H} \right)}
\label{E:phi_iso}
,\end{equation}

        where $\nu$ is the refractivity, which decreases exponentially with \textit{r}, and $r$ is the minimum distance \revis{of} the considered ray path from the center of Venus. This quantity is related to the gas number density \textit{n} by $\nu\ = K n$, where K is the specific refractivity. \textit{H} is the scale height of the atmosphere. 
        
        The factor by which the image of an element of the photosphere is shrunk by refraction is given by
        \begin{equation}
        \phi = \frac{1}
        {1+\frac{D}{d}\left(
        \frac{\partial \omega}{\partial r} 
        \right)} 
        \label{E:singlelayer}
        ,\end{equation}
        
in which $d = 1 + \frac{D}{D'}$, with \textit{D'} and \textit{D} representing the distance of Venus from the Sun and Earth, respectively. At the transit epoch, $D = 0.288703$~AU and $D' = 0.726023$~AU. 
        
        $\phi$ is also the ratio between the flux received by the observer from that element and its flux before refraction, if the atmosphere is completely transparent. 
        
By defining the conventional distance $r_{1/2}$ as the half occultation radius (measured from the planet center) at which $\phi = 0.5$, the following equation is derived:
                
\begin{equation}
\frac{1}{d} \left(\frac{1}{\phi(r)}-1 \right) + \log\left(\frac{1}{\phi(r)}-1 \right) = \frac{ r_{1/2} - r}{H}
\label{E:Baumcode}
.\end{equation}

        We note that Eq.~\ref{E:Baumcode} is valid in the range of $r$ (distance from the center of Venus) spanning from an altitude where the atmosphere is opaque (optical thickness $\tau >> 1$) to the limit at which the refracted light comes from the solar limb (smaller deviations do not reach the observer). While we assume that the inferior limit is constant, the upper one depends on the geometry and, for a given location at the planet limb, changes for different $f$ values. 
        
        The total flux of the aureole will be the integral of the refracted light passing at that distance range from the center
of Venus, that is,

\begin{equation}
F = \int_{r_{min}}^{r_{max}} S_\odot(r)\phi(r)\tau(r)\ l\ dr
\label{eq:BCflux}
,\end{equation}

where $S_\odot(r)$ is the flux emitted by an element of solar photosphere of size $l$, passing at a minimum distance $r$ from the center of Venus. The function $\tau(r)$ represents an absorption factor that can be included in the integration to reflect the vertical structure of aerosols in the upper haze \citep{Wilquet-et-al-2009, Wilquet-et-al-2012} as detailed in Sect.~\ref{S:boundaryc}. 

\begin{figure}
\includegraphics[width=\linewidth]{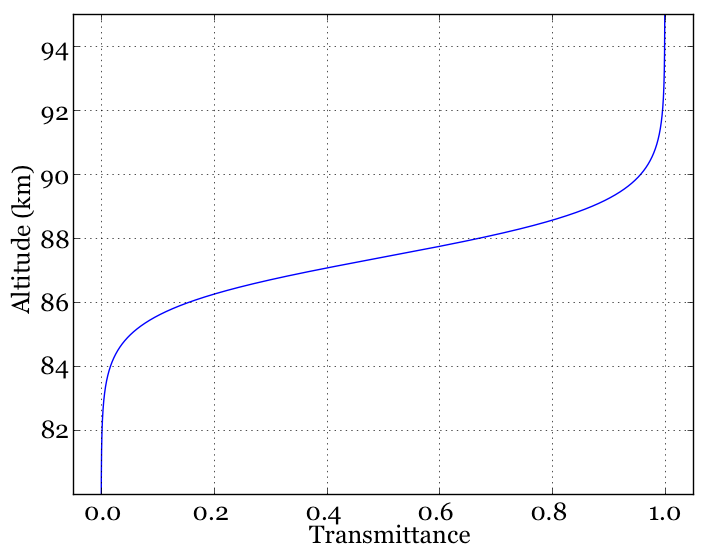}
\caption{Transmission function (Eq.\ref{eq:absorption})
through the atmosphere of Venus adopted by our model. The scale height of the aerosol for $\tau = 1$ at r = 87.4 km, corresponding to $k_{\tau} = 0.6$ km and $H_{\tau} = 4.8$~km, is based on \citet{Wilquet-et-al-2009}.}
\label{absorption_factor}
\end{figure}

To model the aureole, we describe the brightness of the solar disk by a simple limb-darkening function, which yields  $S_\odot(r)$ \citep{Hestroffer-Magnan-1998}. 

In this model, the free parameters are $H$ and $\Delta r = r_{1/2}-r_{\tau}$, where $r_{\tau}$ is the distance from the planet center at which $\tau(r_{\tau}) = 1$.

\subsection{Multilayer model (model 2)}

The isothermal approach, which provides only averaged quantities on the altitudes that generate the aureole, does not appear to be well suited for reproducing \revis{the} observed temporal brightness variations in the aureole flux at $f > 0.5$ (Sect. 3). We therefore implemented a ray–tracing approach considering a multilayered atmosphere, in which each layer is described by its refractive properties, called model 2 in this paper. In our case, the vertical distribution of the refractive index $N(r)$, sampled by a number of  $n$ layers, is the unique physical quantity determining the trajectory of a light beam through the atmosphere. 

This model is entirely equivalent to those used for stellar occultations \citep{Ververka-Wasserman-1973, Elliot-1992, Elliot-2003} and is based on the computation of the total refraction angle resulting from discretizing the path integral of the smoothly varying direction of propagation: 

\begin{equation}
   \theta(r) = \int_{-\infty}^{+\infty} \frac{r}{r'} \frac{d}{dr'} \ln N(r')\ dx
   \label{eq:angle_deviation}
,\end{equation}

in which $\theta(r)$ is the deviation angle of a light beam passing at a minimal distance $r$ from the planet center. $r'>r$ represents the atmospheric altitudes crossed by the light ray above $r$, $N(r')$ is the vertical refractive index profile, and $dx$ is the integration path along the ray propagation.

It has been shown \citep{Elliot-1992} that the corresponding geometric attenuation of the light beam that is due to refraction is equivalent to the integral of several isothermal layers, each one contributing as in Eq.~\ref{E:singlelayer}, that is,

\begin{equation}
   \phi = \int_{-\infty}^{+\infty} \frac{1}{1+ \frac{D}{d}\frac{d\theta(r)}{dr}} dx.
   \label{eq:compute_flux}
\end{equation}

\subsection{Upper haze boundary condition } 
\label{S:boundaryc}

In all the numerical models adopted and compared to SDO/HMI data in Sects. 3.1-3.3, we took into account the geometry of the transit, which evolves with time, to compute sunlight refraction from the source (the solar photosphere) to the observer (placed on Earth or on a space satellite as in the case of SDO). 
    
    The integrals needed to compute the contribution of each atmospheric layer to the aureole (either Eq. \ref{eq:BCflux} or \ref{eq:compute_flux}) were computed over an appropriate altitude range, from layers for which the optical thickness is $<<$1 up to $\sim$140~km, that is, above the region where the aureole is produced. 

To introduce a more realistic transition between the transparent atmosphere and the opaque cloud layers, we introduced a simple optical thickness variation with the altitude $z$, represented by the function

\begin{equation}
   \tau(z) = 0.5+0.5\ \tanh\left[k_{\tau}\ (r-r_{cloud})\right]
   \label{eq:absorption}
,\end{equation}

in which $r_{cloud}$ is the radius of the $\tau$=1 level (Fig.~\ref{absorption_factor}). The parameter $k_{\tau}=0.6\ (km)$ is chosen in such a way that the shape of the variation fits the profiles for aerosol absorption obtained by \citet{Wilquet-et-al-2009}. The scale height of the aerosols is $H_{\tau}=\frac{2.88\ (km)}{k_{\tau}}=4.8$~km.

\section{Results of aureole photometry vs modeling}
\subsection{Isothermal mesosphere (model 1a)}

In the isothermal approach of model 1, the aureole brightness for a given latitude is uniquely determined by the value of the physical scale height ($H$) and by the layer thickness: $\Delta r = r_{1/2} - r_{cloud}$ \citep{Tanga-etal-2012}.

The best fit to the isothermal model is obtained by a mixed \textit{\textup{\textit{Genetic}}} \citep{Holland-1975, Goldberg-1989, Davis-1991, Beasley-1993a, Beasley-1993b, Michalewicz-1994} and Markov chain Monte Carlo
(MCMC) approach \citep{Metropolis-et-al-1953, Hastings-1970, Numerical-Recipes-2007}. 
 

The \textit{Genetic} algorithm is the computation of the best solution between two vectors of possible \textit{H} and $\Delta r$ values, evaluated on the base of the least-squares residuals between the computed flux and the observations. 
The first generation spans a wide range in the parameter space, from 0 to 50 km for \textit{H} and from 0 to 40 km for $\Delta r$. Each additional generation selects the best solutions and narrows the search on a more restrictive set of parameters.
The third generation of the \textit{Genetic} algorithm was used to initiate the MCMC code, which is iterated a number of times sufficient to reach a non-linear least-square minimization condition.

\revis{The aureole flux values that we need to model span more than 2 orders of magnitude, but in terms of physical interest, all brightness levels are equally relevant, including the fainter aureole associated with refraction by deeper atmospheric levels. For this reason, all our fits were computed on the logarithm of the flux, not the flux itself.}

\begin{figure}
\includegraphics[width=\linewidth]{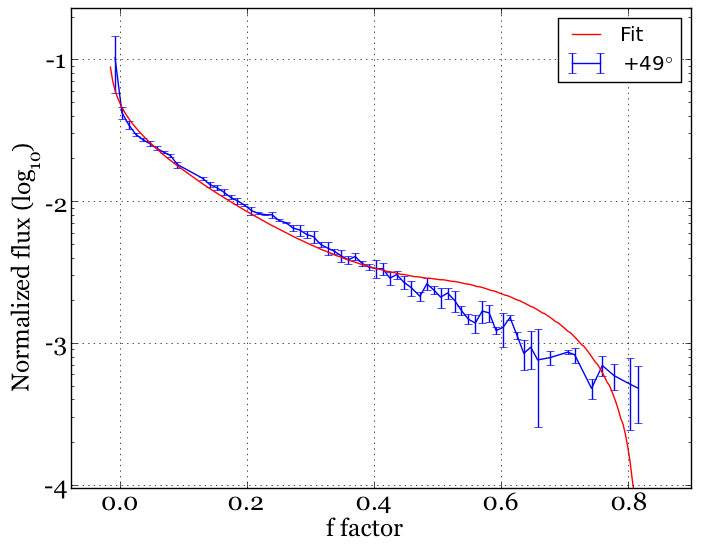}
\caption{Best fit of the aureole light curve \revis{from SDO/HMI} at +49\textdegree (morning terminator) obtained with the single-layer, isothermal model. The flux is the same than in Fig.~\ref{F:lightcurve}, binned over ten consecutive points.         
Parameters are $H=16.3$ km, $r_{cloud} = 94$ km and $r_{1/2}=96$ km. We obtain a significant flux excess in the model at $f > 0.5$ when the Venus limb is more distant from the solar limb, i.e., when higher altitude refractive layers are probed. See Table~\ref{T:res_BC} for the model parameters.}
\label{F:res_BC}
\end{figure}

The results obtained from our photometry at a latitude +49\textdegree are illustrated in Table~\ref{T:res_BC} and Fig.~\ref{F:res_BC}.
\revis{Error bars for the parameters are estimated by searching for the largest parameter variation that fits the standard deviation of the measurements.}

\begin{table}    
    \centering
    \begin{tabular}{c c c c}
    \hline \multicolumn{4}{c}{One-layer model} \\ \hline  
    \hline Latitude & $H$ (km) & $r_{1/2}$(km) & Altitude (km) \\ \hline     
   $+49$\textdegree \ & $16.3\pm0.7$ & $96\pm1$ & $190\pm1$   \\ 
    \hline                                             
    \end{tabular}
    \caption{Result of the model fit for the entire aureole flux variation.}   
    \label{T:res_BC}           
\end{table}


While model 1 appears to reproduce the light curve for $f<0.4$
very closely, the faint aureole appears to be systematically  overestimated up to $f\sim0.7$, where an abrupt cutoff occurs, resulting in a reduced chi square $\tilde{\chi}^2$ of 20. As at the cutoff the signal falls at noise level, the agreement of the model could be considered qualitatively acceptable, but the physical parameters thus determined do not appear to be realistic. In particular the high value of $H=16.3$~km disagrees strongly with other determinations. We can compare it to the typical scale height measured on the SOIR density profile (Fig. \ref{F:dens_SOIR}-top), which is 3-4 times smaller. 

This finding apparently indicates that the isothermal approach is not entirely appropriate. The behavior of the model at $f>0.4$, corresponding to the sunlight passing at lower altitudes, also
suggests a change in the trend of the refractive properties with altitude. As refraction is related to density, the layered scale height distribution obtained by SOIR could play a significant  role in the formation of the aureole. For this reason, we decided to adopt this three-layer structure as an intermediate step toward a more complex modeling. 
  
\subsection{Three isothermal layers (model 1b) }

A variant of model 1 consists of a sliced analysis of the light curve. In fact, the geometry of the refraction is such that for increasing $f$ , only light rays passing deeper in the atmosphere can reach the observer. By considering the faint end of each light curve portion, only the atmosphere closer to the opaque cloud top, where deviation by refraction is maximum, contributes to the aureole. 

\begin{figure}
\includegraphics[width=\linewidth]{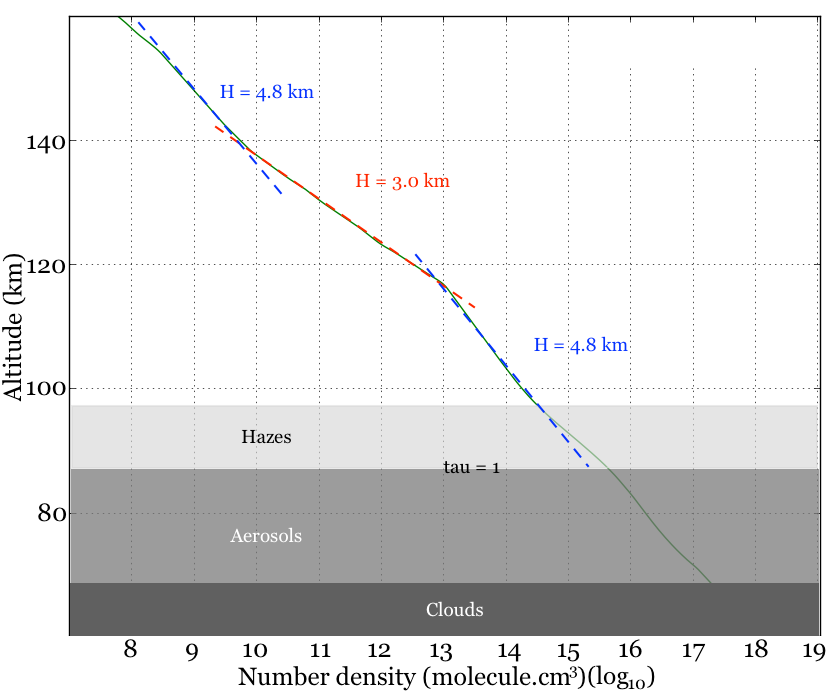}
\includegraphics[width=\linewidth]{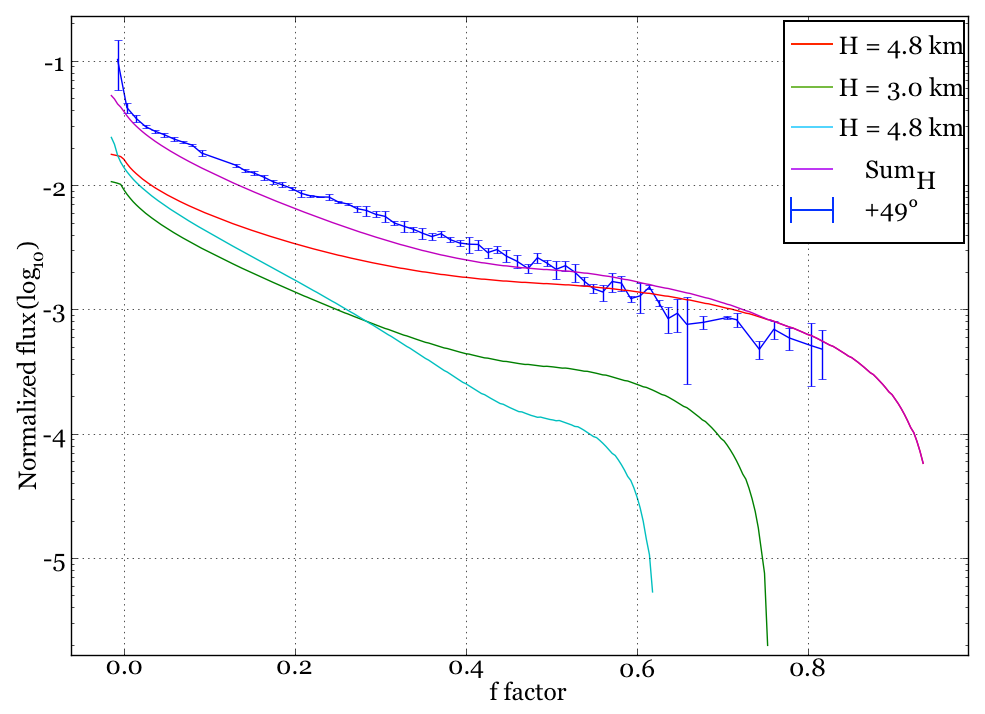}
\caption{Top: molecular density measured by SOIR. \revis{We consider three main layers above the cloud top} ($r_{cloud} \sim 94$ km). The slope of a linear fit over three segments provides three scale heights, $H_1 = 4.8$ km for altitudes below 116 km, $H_2 = 3.0$ km (116 to 135 km), and $H_3 = 4.8$ km (135 to 160 km). Bottom: results of the three-layer modeling (best fit). See Table \ref{T:results} for the model parameters. \revis{The purple (upper) curve represents the summation of the fluxes from each of the three layers, here represented in cyan, green, and red. The blue data points are the measurements, as reported in Fig.~\ref{F:res_BC}.} }
\label{F:dens_SOIR}
\end{figure}

We assumed the SOIR measurement as an input to model 1b. \revis{As shown by the piecewise linear fit in the top panel of Fig.~\ref{F:dens_SOIR}, we considered that} a first layer exists at low altitudes and corresponds to $H=4.8$~km, mostly contributing at the faint end of the light curve $0.4<f<0.6$. The highest level (with the same scale height) should contribute only to the brightest peak ($f<0.1$), while the intermediate level \revis{($H=3$~km) should be relevant in the intermediate portion of the light curve}. Each of the three layers should then replicate the behavior of the isothermal model, within the corresponding altitude range given by the vertical profile of SOIR. As the altitude ranges and the scale heights are provided by SOIR, the only free parameters in this model are the three values of $\Delta r$. 

The results of the fit, computed by the same method as introduced above, are presented in the bottom panel of Fig.~\ref{F:dens_SOIR} and Table~\ref{T:results}\revis{, as a result of the application of the direct flux modeling (Eq.~\ref{eq:BCflux})}.
In the isothermal inverse model and the triple-layer model, the scale height of the aerosols was $H_{\tau}=5.8$~km, the constant was $k_{\tau}=0.5$~(km), and the altitude of the $\tau =1$ was $r_{\tau}=80.0$~km. This altitude was chosen following  \citet{Wilquet-et-al-2012}. 

 It is interesting to note that, as expected, all the three layers contribute to the aureole for $f<0.3$, while the deepest layer dominates for $f>0.4$. However, the final result is not yet fully satisfactory as the flux is in general underestimated (by a factor up to $\sim$2) except for $f>0.5$,\revis{ yielding $\tilde{\chi}^2\approx100$}.

These results seem suggest that a more complex model that is capable of reproducing more details of the vertical density profile
might fit the observations better. 
 
\begin{table*}    
    \centering
    \begin{tabular}{c | c c c | c c c}
    \hline \multicolumn{7}{c}{Parameters and results of the three--layer model} \\ \hline  
    \hline Layer (km) & $\mu_{SOIR}(r)$ & $T_{SOIR} (K)$ & $g(r) (m\ s^{-2})$ & $H$ (km) & $\Delta r$ (km) & $T (K)$ \\ 
    \hline     
     94-116 & 43.22 & 243$\pm$12 &  8.56 & 4.8$\pm$0.5 & 38.0$\pm$2.5 & 214$\pm$10  \\
    116-135 & 42.12 & 141$\pm$12 &  8.51 & 3.0$\pm$0.5 & 47.1$\pm$2.5 & 129$\pm$10  \\ 
    135-160 & 35.55 & 209$\pm$24 &  8.45 & 4.8$\pm$0.5 & 73.0$\pm$2.5   & 173$\pm$10  \\ 
   
    \hline                                             
    \end{tabular}
    \caption{Results of the three-layer model at the latitude +49\textdegree (morning terminator). The molecular weight and the $T_{SOIR}$ temperature at the average altitude of each layer obtained from the data processing of orbit 2238 are reported together with the value of the gravity acceleration g(r).}   
    \label{T:results}           
\end{table*}

We can also compare the temperatures that are represented by the fully resolved vertical profile to our three--layer model. We use the equation

\begin{equation}
   T = \frac{\mu(z)g(z)H(z)}{R}
   \label{eq:temperature}
,\end{equation}

where R is the ideal gas constant and $\mu$ the mean molecular weight measured by SOIR. Gravity g(z) depends on altitude. For all the three layers we computed the corresponding quantities at their average altitude. The temperature values in the three-layer models do not exactly correspond to those obtained when the full profile of variations is taken into account, which additionally underlines the evidence that local fluctuations in the atmospheric scale height can be relevant. 


    
       

\subsection{Multilayer model (model 2)}
        
By representing the atmosphere over several layers, whose vertical extent is much smaller than the typical scale height, we wish to test whether a better modeling of the photometry, relative to the three-layers approach, can be obtained. In turn, we will be able to investigate the sensitivity of the aureole photometry to small details of the vertical temperature profile.
        
To integrate Eq.~\ref{eq:compute_flux} we discretized it on a set of $m$ atmospheric layers of equal thickness. In our case $m$=400 and the thickness $\delta$r=400~m. Each layer was associated with a different refractivity $\nu(r)$. By considering a pure CO$_2$ atmosphere, we computed the refractivity as $\nu(r) = K\ n(r)$, where the specific refractivity of CO$_2$ is $K=$1.67$\times$10$^{-29}$~m$^3$~molecule$^{-1}$ \citep{CO2}. $n(r)$ is the number density provided by SOIR following the approach in \citet{Mahieux-et-al-2015a}. 

The core of the computation is the application of Eq.~\ref{eq:compute_flux} at all layers; this provides the total refraction angle and the associated attenuation. From the refraction angle, a light ray is traced back from the observer toward the source on a plane containing the observer, the center of Venus, and the point of the terminator where refraction must be analyzed (at +49\textdegree in our case). When the ray falls on the solar photosphere, the corresponding flux contribution (weighted by $\phi$) is considered. The aerosol absorption factor \ref{eq:absorption} is also used to model the transition between the transparent and the opaque atmosphere.

By adding the contribution of each layer, we obtained the aureole theoretical brightness. The computation was then repeated for each $f$ to reconstruct the full light curve, to be compared with SDO/HMI observations.



\begin{figure}
\includegraphics[width=\linewidth]{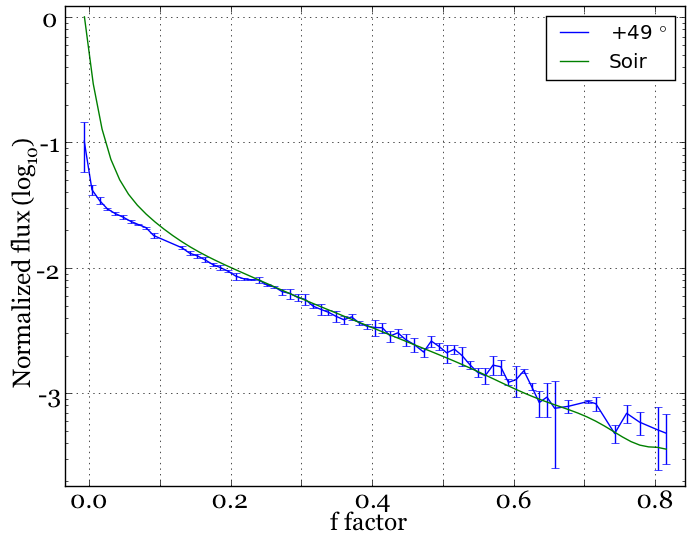}
\includegraphics[width=\linewidth]{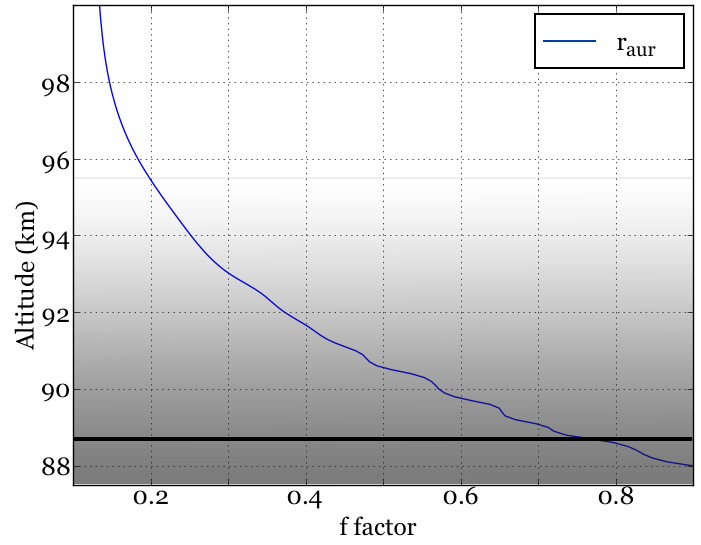}
\caption{\revis{Top panel:} best fit obtained with the vertical density profile of SOIR for the latitude +49$^{\circ}$ and an aerosol scale height $H = 4.8$ km using the multilayer refractive model. \revis{Bottom panel:} altitude of the highest layer probed by the aureole as a function of f (blue solid line). The maximum altitude corresponds to the tangent limb geometry at which the source of the light ray reaching the observer is the solar limb. Above this altitude, the deviation by refraction is too small to deflect sunlight toward the observer at +49$^{\circ}$. The gray gradient represents the scale height of the aerosols (H = 4.8 km) \revis{used} in model 2. The dark line represents $\tau = 1$.}
        \label{F:multi-fit}
\end{figure}

        The result of this procedure is shown in Fig.~\ref{F:multi-fit}, where the green curve shows the predicted flux obtained by the direct model, and the blue curve represents the SDO/HMI data. The fit agrees remarkably well for $f>0.08$, where the general slope is perfectly reproduced. 
        
        The cloud altitude and the scale height of the aerosols are obtained iteratively by a least-squares minimization. 
The resulting values are $r_{\tau} = 89.0$~km and $r_{\tau}=4.8$~km, respectively. 
        
\revis{We obtain here the best fit from all the models with $\tilde{\chi}^2\approx9$, a value affected by the largely over-estimated aureole brightness for $f<0.08$}. Although the measured flux at this portion of the light curve might be marginally contaminated by the proximity of the solar limb, which makes the measurement rather delicate, our impression is that the difference is real and can probably be ascribed to the assumption of a pure CO$_2$ atmosphere. Figure~\ref{F:multi-fit} shows the altitude of the highest layer contributing to the aureole for each $f$ value. For very low $f$, the quasi--alignment of the solar limb with the refraction point on the Venus terminator, and with the observer, corresponds to very low refraction angles, that is, to high atmospheric levels. 
        
As the aureole reaches an altitude $z\sim 100~km$ from the planet surface, we can assume that fractionation starts to play a role, and other species different from CO$_2$ \citep{Bertaux-2007} need to be taken into account in the computation. These species, like H$_2$O HDO \citep{Federova-2008}, SO \citep{Bertaux-2007}, SO$_2$ \citep{Bertaux-2007, Belyaev-2012, Mahieux-et-al-2015a}, CO, O, He, N, and N$_2$ \citep{Vandaele-2016} or HCl / HF \citep{Mahieux-et-al-2015b} are present in the atmosphere and have a lower refractivity than CO$_2$. These could decrease the contribution of the highest atmospheric levels to the aureole brightness.
        

\section{Conclusion and perspectives}

A new procedure for the photometry of the aureole was implemented
that provides accurate measurements of the elusive brightness
of the aureole all along the Venus terminator. The time resolution of the SDO images analyzed in this paper is much higher than the one available in 2004, and the photometry is of much better quality, mainly because these observations were obtained from space.

For the first time we were able to compare the vertical density profile obtained by SOIR to remote observations of the aureole, showing that the SOIR profile is capable of reproducing the general features. In the process we showed that the measured aureole flux is sensitive to details in the vertical profile.

We compared three different approaches that can be used to model the aureole brightness. The first is based on a transparent isothermal atmosphere as described by \citet{Baum-Code-1953}. This approach was adopted to analyze the much less accurate data of the transit in 2004 \citep{Tanga-etal-2012}. 

The second approach consists of an extension of this model to three isothermal layers by using the information provided by the SOIR experiment at +49\textdegree, from observations secured by the Venus Express during the solar transit event itself. The SOIR vertical density profile clearly exhibits three ranges in which, at first order, the temperature can be considered as constant.

The final attempt adopted a multilayer model with a layer thickness much smaller than any physical scale height. The full resolution of the vertical density profile by SOIR was adopted in this model.

A comparison of the three methods showed that only the last model reproduces the trend of the aureole light curve with reasonable accuracy. This finding further indicates the sensitivity of the aureole to subtle details in the vertical density profile. As the model is based on the direct application of the SOIR--derived profile, our result is also an independent confirmation, from remote observations, of the results obtained by SOIR. 

The only free parameter of the multilayer approach, the altitude of the $\tau=1$ level, has a value compatible with other determinations of the upper cloud deck limit altitude. 

Our model adopts a simplified vertical distribution of aerosols that can be further improved or tested against more recent SOIR data. However, we find no clear discrepancy that can be attributed to a lack of detail in the aerosol distribution.

After assessing the reliability of our photometry and modeling against the SOIR data, we will explore in following publications the vertical density and temperature profiles at other latitudes. Additional developments are due, by the implementation of an inverse model of Eq.~\ref{eq:compute_flux}, to be illustrated in a forthcoming publication.
   
\begin{acknowledgements}

This research is supported by the European Commission Framework Program FP7 under Grant Agreement 606798 (Project EuroVenus). We credit the National Aeronautics and Space Administration (NASA) and the HMI science team for providing the data. TW acknowledges University of Versailles-St-Quentin, CNES VEx-SI program and France's Programme National de Plan\'etologie.

\end{acknowledgements}

\bibliographystyle{aa}
\bibliography{refbib}

\begin{thebibliography}{34}
\expandafter\ifx\csname natexlab\endcsname\relax\def\natexlab#1{#1}\fi

\bibitem[{{Baum} \& {Code}(1953)}]{Baum-Code-1953}
{Baum}, W.~A. \& {Code}, A.~D. 1953, AJ, 58, 108

\bibitem[{Beasley {et~al.}(1993{\natexlab{a}})Beasley, Bull, \&
  Martin}]{Beasley-1993a}
Beasley, D., Bull, D., \& Martin, R. 1993{\natexlab{a}}, An Overview of Genetic
  Algorithms:Part I, Fundamentals

\bibitem[{Beasley {et~al.}(1993{\natexlab{b}})Beasley, Bull, \&
  Martin}]{Beasley-1993b}
Beasley, D., Bull, D.~R., \& Martin, R.~R. 1993{\natexlab{b}}, An Overview of
  Genetic Algorithms: Part 2, Research Topics

\bibitem[{Belyaev {et~al.}(2012)Belyaev, Montmessin, Bertaux, Mahieux,
  Fedorova, Korablev, Marcq, Yung, \& Zhang}]{Belyaev-2012}
Belyaev, D.~A., Montmessin, F., Bertaux, J.-L., {et~al.} 2012, Icarus, 217, 740
  , advances in Venus Science

\bibitem[{Bertaux {et~al.}(2007)Bertaux, Nevejans, Korablev, Villard,
  Quémerais, Neefs, Montmessin, Leblanc, Dubois, Dimarellis, Hauchecorne,
  Lefèvre, Rannou, Chaufray, Cabane, Cernogora, Souchon, Semelin, Reberac,
  Ransbeek, Berkenbosch, Clairquin, Muller, Forget, Hourdin, Talagrand, Rodin,
  Fedorova, Stepanov, Vinogradov, Kiselev, Kalinnikov, Durry, Sandel, Stern, \&
  Gérard}]{Bertaux-2007}
Bertaux, J.-L., Nevejans, D., Korablev, O., {et~al.} 2007, Planetary and Space
  Science, 55, 1673 , the Planet Venus and the Venus Express Mission, Part 2

\bibitem[{{Bideau-Mehu} {et~al.}(1973){Bideau-Mehu}, {Guern}, {Abjean}, \&
  {Johannin-Gilles}}]{CO2}
{Bideau-Mehu}, A., {Guern}, Y., {Abjean}, R., \& {Johannin-Gilles}, A. 1973,
  {Opt. Commun.}, 9, 432

\bibitem[{Clancy {et~al.}(2015)Clancy, Sandor, \& Hoge}]{Clancy-et-al-2015}
Clancy, R.~T., Sandor, B.~J., \& Hoge, J. 2015, Icarus, 254, 233

\bibitem[{Davis(1991)}]{Davis-1991}
Davis, L. 1991, Handbook of Genetic Algorithms (New York: Van Nostrand
  Reinhold)

\bibitem[{{Ehrenreich} {et~al.}(2012){Ehrenreich}, {Vidal-Madjar}, {Widemann},
  {Gronoff}, {Tanga}, {Barth{\'e}lemy}, {Lilensten}, {Lecavelier Des Etangs},
  \& {Arnold}}]{Ehrenreich2012}
{Ehrenreich}, D., {Vidal-Madjar}, A., {Widemann}, T., {et~al.} 2012, \aap, 537,
  L2

\bibitem[{{Elliot} {et~al.}(2003){Elliot}, {Person}, \& {Qu}}]{Elliot-2003}
{Elliot}, J.~L., {Person}, M.~J., \& {Qu}, S. 2003, \aj, 126, 1041

\bibitem[{{Elliot} \& {Young}(1992)}]{Elliot-1992}
{Elliot}, J.~L. \& {Young}, L.~A. 1992, \aj, 103, 991

\bibitem[{Fedorova {et~al.}(2008)Fedorova, Korablev, Vandaele, Bertaux,
  Belyaev, Mahieux, Neefs, Wilquet, Drummond, Montmessin, \&
  Villard}]{Federova-2008}
Fedorova, A., Korablev, O., Vandaele, A.-C., {et~al.} 2008, Journal of
  Geophysical Research: Planets, 113, n/a, e00B22

\bibitem[{Goldberg(1989)}]{Goldberg-1989}
Goldberg, D.~E. 1989, Genetic Algorithms in Search, Optimization, and Machine
  Learning (Reading, MA: Addison--Wesley)

\bibitem[{Hastings(1970)}]{Hastings-1970}
Hastings, W.~K. 1970, Biometrika, 57, 97

\bibitem[{{Hestroffer} \& {Magnan}(1998)}]{Hestroffer-Magnan-1998}
{Hestroffer}, D. \& {Magnan}, C. 1998, \aap, 333, 338

\bibitem[{Holland(1975)}]{Holland-1975}
Holland, J. 1975, Adaptation in natural and artificial systems : an
  introductory analysis with applications to biology, control, and artificial
  intelligence, 1st edn. (Cambridge Mass.: {MIT} Press)

\bibitem[{{Jaeggli} {et~al.}(2013){Jaeggli}, {Reardon}, {Pasachoff},
  {Schneider}, {Widemann}, \& {Tanga}}]{Jaeggli-et-al-2013}
{Jaeggli}, S.~A., {Reardon}, K.~P., {Pasachoff}, J.~M., {et~al.} 2013, in
  AAS/Solar Physics Division Meeting, Vol.~44, AAS/Solar Physics Division
  Meeting, 100.150

\bibitem[{Link(1969)}]{Link-1969}
Link, F. 1969, Eclipse phenomena in Astronomy (Berlin: Springer-Verlag Berlin,
  Inc.)

\bibitem[{Mahieux {et~al.}(2015{\natexlab{a}})Mahieux, Vandaele, Bougher,
  Drummond, Robert, Wilquet, Chamberlain, Piccialli, Montmessin, Tellmann,
  Pätzold, Häusler, \& Bertaux}]{Mahieux-et-al-2015a}
Mahieux, A., Vandaele, A., Bougher, S., {et~al.} 2015{\natexlab{a}}, Planetary
  and Space Science, 113–114, 309 , sI:Exploration of Venus

\bibitem[{Mahieux {et~al.}(2015{\natexlab{b}})Mahieux, Vandaele, Robert,
  Wilquet, Drummond, Valverde, Puertas, Funke, \&
  Bertaux}]{Mahieux-et-al-2015b}
Mahieux, A., Vandaele, A., Robert, S., {et~al.} 2015{\natexlab{b}}, Planetary
  and Space Science, 113–114, 347 , sI:Exploration of Venus

\bibitem[{{Metropolis} {et~al.}(1953){Metropolis}, {Rosenbluth}, {Rosenbluth},
  {Teller}, \& {Teller}}]{Metropolis-et-al-1953}
{Metropolis}, N., {Rosenbluth}, A.~W., {Rosenbluth}, M.~N., {Teller}, A.~H., \&
  {Teller}, E. 1953, \jcp, 21, 1087

\bibitem[{Michalewicz(1994)}]{Michalewicz-1994}
Michalewicz, Z. 1994, Genetic algorithms + data structures = evolution programs
  (2nd, extended ed.) (New York, NY, USA: Springer-Verlag New York, Inc.)

\bibitem[{{Pasachoff} {et~al.}(2011){Pasachoff}, {Schneider}, \&
  {Widemann}}]{Pasachoff-etal-2011}
{Pasachoff}, J.~M., {Schneider}, G., \& {Widemann}, T. 2011, \aj, 141, 112

\bibitem[{Press {et~al.}(2007)Press, Teukolsky, Vetterling, \&
  Flannery}]{Numerical-Recipes-2007}
Press, W.~H., Teukolsky, S.~A., Vetterling, W.~T., \& Flannery, B.~P. 2007,
  Numerical Recipes 3rd Edition: The Art of Scientific Computing, 3rd edn. (New
  York, NY, USA: Cambridge University Press)

\bibitem[{{Reale} {et~al.}(2015){Reale}, {Gambino}, {Micela}, {Maggio},
  {Widemann}, \& {Piccioni}}]{Reale-et-al-2015}
{Reale}, F., {Gambino}, A.~F., {Micela}, G., {et~al.} 2015, Nature
  Communications, 6, 7563

\bibitem[{{Schou} {et~al.}(2012){Schou}, {Scherrer}, {Bush}, {Wachter},
  {Couvidat}, {Rabello-Soares}, {Bogart}, {Hoeksema}, {Liu}, {Duvall}, {Akin},
  {Allard}, {Miles}, {Rairden}, {Shine}, {Tarbell}, {Title}, {Wolfson},
  {Elmore}, {Norton}, \& {Tomczyk}}]{Schou-et-al-2012}
{Schou}, J., {Scherrer}, P.~H., {Bush}, R.~I., {et~al.} 2012, Solphys, 275, 229

\bibitem[{{Tanga} {et~al.}(2012){Tanga}, {Widemann}, {Sicardy}, {Pasachoff},
  {Arnaud}, {Comolli}, {Rondi}, {Rondi}, \& {S{\"u}tterlin}}]{Tanga-etal-2012}
{Tanga}, P., {Widemann}, T., {Sicardy}, B., {et~al.} 2012, Icarus, 218, 207

\bibitem[{Vandaele {et~al.}(2016)Vandaele, Chamberlain, Mahieux, Ristic,
  Robert, Thomas, Trompet, Wilquet, Belyaev, Fedorova, Korablev, \&
  Bertaux}]{Vandaele-2016}
Vandaele, A., Chamberlain, S., Mahieux, A., {et~al.} 2016, Advances in Space
  Research, 57, 443

\bibitem[{{Vandaele} {et~al.}(2008){Vandaele}, {De Mazi{\`e}re}, {Drummond},
  {Mahieux}, {Neefs}, {Wilquet}, {Korablev}, {Fedorova}, {Belyaev},
  {Montmessin}, \& {Bertaux}}]{Vandaele-et-al-2008}
{Vandaele}, A.~C., {De Mazi{\`e}re}, M., {Drummond}, R., {et~al.} 2008, Journal
  of Geophysical Research (Planets), 113, 0

\bibitem[{{Wasserman} \& {Veverka}(1973)}]{Ververka-Wasserman-1973}
{Wasserman}, L. \& {Veverka}, J. 1973, in Bulletin of the American Astronomical
  Society, Vol.~5, Bulletin of the American Astronomical Society, 289

\bibitem[{{Widemann} {et~al.}(2012){Widemann}, {Tanga}, {Reardon}, {Limaye},
  {Wilson}, {Vandaele}, {Wilquet}, {Mahieux}, {Robert}, {Pasachoff}, \&
  {Schneider}}]{Widemann-et-al-2012}
{Widemann}, T., {Tanga}, P., {Reardon}, K.~P., {et~al.} 2012, in AAS/Division
  for Planetary Sciences Meeting Abstracts, Vol.~44, AAS/Division for Planetary
  Sciences Meeting Abstracts, 508.08

\bibitem[{Wilquet {et~al.}(2012)Wilquet, Drummond, Mahieux, Robert, Vandaele,
  \& Bertaux}]{Wilquet-et-al-2012}
Wilquet, V., Drummond, R., Mahieux, A., {et~al.} 2012, Icarus, 217, 875 ,
  advances in Venus Science

\bibitem[{Wilquet {et~al.}(2009)Wilquet, Fedorova, Montmessin, Drummond,
  Mahieux, Vandaele, Villard, Korablev, \& Bertaux}]{Wilquet-et-al-2009}
Wilquet, V., Fedorova, A., Montmessin, F., {et~al.} 2009, Journal of
  Geophysical Research: Planets, 114, n/a, e00B42

\bibitem[{{Wilson} {et~al.}(2012){Wilson}, {Perez-Ayucar}, {Markiewicz},
  {Vandaele}, {Mahieux}, \& {Bertaux}}]{Wilson-et-al-2012}
{Wilson}, C.~F., {Perez-Ayucar}, M., {Markiewicz}, W.~J., {et~al.} 2012, in
  European Planetary Science Congress 2012, EPSC2012--913

\end{thebibliography}

\end{document}